\documentclass[aps,preprint]{revtex4}%
\usepackage{amsfonts}
\usepackage{amsmath}
\usepackage{amssymb}
\usepackage{graphicx}%
\begin{document}
\title{Fading entanglement near equilibrium state}
\author{Gregory B. Furman, Victor M. Meerovich, and Vladimir L. Sokolovsky}
\affiliation{Physics Department, Ben Gurion University of the Negev, Beer Sheva, 84105 Israel}

\begin{abstract}
It was recently shown that entanglement in quantum systems being in a
non-equilibrium state can appear at much higher temperatures than in an
equilibrium state. However, any system is subject to the natural relaxation
process establishing equilibrium. The work deals with the numerical study of
entanglement dynamics in a dipolar coupled spin-1/2 system under the
transition from a non-equilibrium state to an equilibrium state. The spin
system is characterized by a two-temperature density matrix, and the process
of the establishment of equilibrium is in the equalization of these
temperatures. The method of the non-equilibrium statistical operator is used
to describe the evolution of the system. The process of establishing an
equilibrium state in the homonuclear spin systems at low temperature was first
considered. It was shown that the time dependences of the inverse temperatures
of the spin subsystems are given by a solution of non-linear equations in
contrast to the linear equations in the well-known high temperature
approximation. It was first studied the entanglement dynamics during the
equilibrium state establishment and found that, during establishing
equilibrium, the concurrence changes non-monotonically with time and
temperatures. Entanglement fades long before equilibrium is established in the
system. It was shown that the entanglement dynamics depends strongly on the
ratio of the Zeeman energy to the dipolar energy. At a high ratio, the
concurrence in the system decreases quickly for time about 100 $\mu s$ while,
at a low ratio, establishment of equilibrium and fading entanglement take
prolonged time up to 1 $ms$.

\end{abstract}
\maketitle

\section{Introduction}

Entanglement is a term used in quantum theory to describe the fact that the
particles can be correlated with each other regardless of how far they are
from each other \cite{Benenti2007,Amico2008,Horodecki2009}. Many current
studies \cite{Benenti2007,Amico2008,Horodecki2009} focus on how to harness the
potential of entanglement in the development of quantum cryptography, quantum
communication \cite{Bennett1993}, quantum computation \cite{Bennett2000}, and
quantum metrology \cite{Cappellaro2005,Roos2006}. The growth in interest in
entanglement applications has stimulated intensive qualitative and
quantitative research of entanglement in various physical systems and in both
equilibrium
\cite{Amico2008,Horodecki2009,Amico2004,Doronin2009,FurmanQIP2011,FurmanQIP
2011a,FurmanPLA2012} and non-equilibrium \cite{L. Amico2004A,V.
Subrahmanyam2004,N. Buric2008,G. B. FurmanPRA2008,G. B. FurmanQIP2009,E. B.
Fel'dman2009}, including the pseudo-pure state
\cite{CoryPure,GershenfeldPure,FurmanPure2006} as a initial state
\cite{FurmanPure2009}.

It was shown that quantum systems in thermal equilibrium became entangled, if
the interaction energy between them is larger than thermal energy due to their
coupling to the environment. This condition leads to that entanglement appears
only at very low temperature.

For example, entanglement in dipolar coupled spin system in equilibrium state
is achieved by the application of a low external magnetic field in which the
Zeeman interaction energy is the order of or even less than the dipolar
interaction energy, $\omega_{0}\leq\omega_{d}$ \cite{FurmanQIP2011,FurmanQIP
2011a}. For a proton spin systems, the dipolar energy is $\omega_{d}=15$ kHz.
That results in the temperature of the entanglement appearance of $T\sim
\frac{\omega_{d}}{k_{B}}=0.\allowbreak7\mu K\,\allowbreak$
\cite{Doronin2009,FurmanQIP2011,FurmanQIP 2011a,FurmanPLA2012}. Here $k_{B}$
is the Boltzmann constant.

Recently, it was proved that this relationship between temperature and
entanglement is not justified for systems that are not in thermal equilibrium
\cite{Galve2010}. In contrast to the equilibrium cases, as it was shown that
entanglement in quantum systems being in a non-equilibrium state can appear at
much higher temperatures \cite{Galve2010,V. Vedral2010}. Investigations of
spin systems demonstrate this statement: e.g. a spin system, which is in a
separated state at equilibrium, is irradiated by a radiofrequency field and
goes out from the equilibrium, and entanglement appears in the system
simultaneously \cite{L. Amico2004A,V. Subrahmanyam2004,N. Buric2008,G. B.
FurmanPRA2008,G. B. FurmanQIP2009,E. B. Fel'dman2009}{\Large .}

The present paper deals with the opposite case when the system being initially
in a non-equilibrium state evolves to an equilibrium state. This entanglement
dynamics during the natural process of establishing equilibrium is an actual
problem which, according to our knowledge, does not studied yet. We study
entanglement dynamics in a dipolar coupled spin-1/2 system under the
transition from non-equilibrium state to equilibrium state. We consider a spin
system, which can be divided to two thermodynamic subsystems which are in the
equilibrium state and characterized by two different temperatures (so-called a
quasi-equilibrium state \cite{Goldman M 1970,AbragamGoldman1982}). The process
of establishing equilibrium in the whole system can be represented as
equalization of their temperatures. Such quasi-equilibrium states were
prepared experimentally using various methods \cite{Goldman M
1970,AbragamGoldman1982,J. Jeener1964}. These states can be considered as the
last stage in relaxation process from non-equilibrium state to equilibrium
one. Existence of such states can be explained by using the fundamental
principles of non-equilibrium statistical mechanics first formulated by
Bogolyubov \cite{Bogolyubov1962}. In non-equilibrium statistical mechanics, a
fundamental role is played hierarchy of the relaxation times and reducing in
the number of parameters requested for the description of a non-equilibrium
process. As shown by Bogolyubov, after the lapse of a small interval of time,
of the order of the duration of a collision, all higher distribution functions
are completely determined by a single-particle function, and kinetic equations
can be found for this stage . This means it is possible to use an abbreviated
description of a non-equilibrium state in which the complete distribution
function depends on few parameters.

Bogolyubov's idea about hierarchy of the relaxation times has been widely used
for description of relaxation processes in spin systems \cite{A. Abragam
1961,Redfield,Provotorov1961,Goldman M 1970}. In particular, if the
correlation times are much shorter than the time $T_{2},$ the lifetime of the
free precession signal, and the time $T_{2}$ is shorter then the
characteristic time of an investigated process, then the equilibrium state is
established in parts of the spin system before the whole system reaches the
total equilibrium.

The structure of the paper is as follows: in the next section, we describe the
Hamiltonian for a spin system in an external field at the equilibrium and
quasi-equilibrium states and principale of the Provotorov saturation theory
\cite{Goldman M 1970,Provotorov1961}. According to this theory in a dipolar
coupled spin system in a quasi-equilibrium state, two spin temperatures
$T_{z}$ and $T_{d}$ can be assigned to the Zeeman and dipolar thermodynamic
subsystems \cite{Goldman M 1970,Provotorov1961}. In section 3 establishment of
the equilibrium state is analyzed. Then we consider dynamics of the pairwise
entanglement in a quasi-equilibrium state which is characterized by these two
temperatures. Discussion of the results is given in the final section.
Numerical calculations of evolution of the spin temperature and concurrence
$C_{mn}$ between the $m$-th and $n$-th spins are performed using the software
based on the MatLab package. To demonstrate dependence of the entanglement
dynamics on the relative spin position we will consider three four-spin
systems: linear chain, ring, and rectangle. The choice of these systems is due
to the following reasons: i) these structures exist in real molecules and used
in experiments and ii) as dipole-dipole interactions decrease inversely to the
third power of the distance between the spins, the effect of the system
geometry is manifested more clearly at a small number of spins.

\section{Spin system in quasi-equilibrium state}

We consider the system of identical spin-1/2 particles with the dipole-dipole
interaction (DDI) in an external magnetic field $\vec{H}_{0}$ along the $z$ -
axis. In the equilibrium state the system is characterized by a density matrix%

\begin{equation}
\rho=\frac{\exp\left(  -\beta H\right)  }{Z\left(  \beta\right)  }, \tag{1}%
\end{equation}
with a single parameter $\beta=\frac{1}{k_{B}T}$ denotes the inverse spin
temperature, $T$ is the spin temperature, and $Z\left(  \beta\right)
=Tr\left\{  \exp\left(  -\beta H\right)  \right\}  $ is the partition
function. The Hamiltonian consists of the two parts, $H=$ $H_{Z}+H_{dd}$, the
Zeeman part
\begin{equation}
H_{z}=-\gamma\vec{H}_{0}\sum_{j}\vec{I}_{j} \tag{2}%
\end{equation}
and the dipolar part
\begin{equation}
H_{dd}=-\gamma^{2}\hbar\sum_{jk}\left(  \frac{3}{r_{jk}^{5}}\left(  \vec
{I}_{j}\vec{r}_{jk}\right)  \left(  \vec{I}_{k}\vec{r}_{jk}\right)  -\frac
{1}{r_{jk}^{3}}\vec{I}_{j}\vec{I}_{k}\right)  . \tag{3}%
\end{equation}
Here $\gamma$ is the gyromagnetic ratio, $\vec{I}_{j}$ is the angular momentum
operator of the $j$-th spin ($j$ = $1,2...N$), $\vec{r}_{jk}$ is the radius
vector from the $j$-th to $k$-th spins with the spherical coordinates $r_{jk}%
$, $\theta_{jk}$ and $\varphi_{jk}$ . $\theta_{jk}$ \ is the angle between the
radius vector $\vec{r}_{jk}$ and direction of the external magnetic field.

In low magnetic field, when the Zeeman energy splitting $\omega_{0}=\gamma
H_{0}$, and DDI energy $\omega_{d}=\sqrt{\frac{TrH_{dd}^{2}}{Tr\left(
\sum_{j}I_{zj}\right)  ^{2}}}$ \ are commensurable quantities, the time for
reaching equilibrium state describing by Eq.(1) is determined by $\omega
_{d}^{-1}$ which is of the order of $T_{2}$ \cite{AbragamGoldman1982}.

In high magnetic field, $\omega_{0}>>\omega_{d}$, the time $t$ for reaching
equilibrium becomes very long, $t>>T_{2}$ , due to the fact that the energy
spectra of the Zeeman energy and the DDI energy are very different
\cite{AbragamGoldman1982}.

We will consider the spin system placed in high magnetic field as two
thermodynamic subsystems, the Zeeman and dipole-dipole ones, each of them is
at thermodynamic equilibrium with own spin temperature and between which there
is an energy transfer. This approach is based on the following reasoning.

The total Hamiltonian can be presented as the sum of two terms each commuting
with other , $H_{z}$ and $H_{d}$ $\left(  \left[  H_{z},H_{d}\right]
=0\right)  $, and the term $H_{nd}=H_{dd}-H_{d}$, which does not commute with
both $H_{z}$ and $H_{d}$. $H_{d}$ is the secular part of dipolar Hamiltonian
(3)%
\begin{equation}
H_{d}=-\gamma^{2}\hbar\sum_{jk}r_{jk}^{-3}P_{2}\left(  \cos\theta_{jk}\right)
\left(  3I_{zj}I_{zk}-\vec{I}_{j}\vec{I}_{k}\right)  , \tag{4}%
\end{equation}
where $P_{2}\left(  \cos\theta_{jk}\right)  =\frac{1}{2}\left(  1-3\cos
^{2}\theta_{jk}\right)  $, $I_{\alpha j}$ is the projection of the angular
momentum operator of the $j$-th spin ($j$ = 1, 2 , . . ., N) on the $\alpha$ -
axis $(\alpha=x,y,z)$. At $\omega_{0}>>\omega_{d}$ the effect of the dipolar
interaction is small in comparison with the Zeeman interaction and, in the
first-order perturbation, the secular part $H_{d}$ (4) plays the main role in
establishing equilibrium in each subsystem. This process is characterized by
flip-flop terms in the secular part $H_{d}$ , that ensures the density matrix
in the following approximate form%
\begin{equation}
\rho=\frac{\exp\left(  -\beta_{z}H_{z}-\beta_{d}H_{d}\right)  }{Z\left(
\beta_{z},\beta_{d}\right)  }. \tag{5}%
\end{equation}
where $\beta_{z}=\frac{1}{k_{B}T_{z}}\ $and $\beta_{d}=\frac{1}{k_{B}T_{d}}$
are the inverse Zeeman and dipolar temperatures, respectively, $Z\left(
\beta_{z},\beta_{d}\right)  =Tr\left\{  \exp\left(  -\beta_{z}H_{z}-\beta
_{d}H_{d}\right)  \right\}  $ is the partition function that encodes the
statistical properties of the system in the quasi-equilibrium state.
Separation of the spin system into the Zeeman and dipolar subsystems is
consequence of the fact that the Zeeman
\begin{equation}
\left\langle H_{z}\right\rangle =\omega_{0}\left\langle I_{z}\right\rangle
=-\frac{\partial\ln Z\left(  \beta_{z},\beta_{d}\right)  }{\partial\beta_{z}}
\tag{6}%
\end{equation}
and dipolar energies%
\begin{equation}
\left\langle H_{d}\right\rangle =-\frac{\partial\ln Z\left(  \beta_{z}%
,\beta_{d}\right)  }{\partial\beta_{d}} \tag{7}%
\end{equation}
are separately constants of the motion , while both $\left\langle
H_{z}\right\rangle $ and $\left\langle H_{d}\right\rangle $ depend on both
thermodynamics parameters, $\beta_{z}\ $and $\beta_{d}$
\cite{AbragamGoldman1982}. Results (5-7) have been conformed by experiments
\cite{AbragamGoldman1982,J. Jeener1964}. $\ $

The time for reaching the quasi-equilibrium state described by Eq. (5) is of
the order of $T_{2}$. It is assumed that further relaxation process takes more
time $t>$ $T_{2}$. This process consists in equalization of the temperature
$\beta_{z}^{-1}\ $and $\beta_{d}^{-1}$ and results in establishing of the
equilibrium state which is characterized by density matrix (1) with
$\ \beta_{z}\ $=$\beta_{d}$ =$\beta$.

The approximate form of the density matrix does not include the non-secular
term of the dipolar Hamiltonian, $H_{nd}$ . To take into account this term and
to describe the relaxation at time $t>$ $T_{2}$, we will use below the method
of the non-equilibrium statistical operator \cite{Zubarev1974} .

\section{Establishment of equilibrium}

During the time $T_{2}$ the system loses the quantum coherence features and
achieves a quasi-equilibrium state due to the flip-flop terms including in the
Hamiltonian $H_{d}$. For the spin system at low temperature, the decoherence
time $T_{2}$ typically ranges between nanoseconds and seconds
\cite{DiVincenzo1995}. A typical valuem of $T_{2}$ in a nuclear dipolar
coupled spin system is 100 $\mu s$ \cite{Abragam1978} which is much shorter
than the spin-lattice relaxation times of the Zeeman, $T_{1z}$ and dipolar,
$T_{1d}$ energies. \ $T_{1z}$ and $T_{1d}$ take values in the range from
minutes to hours \cite{AbragamGoldman1982}. We will study spin dynamics at
$T_{2}<t\ll\min\left\{  T_{1z},T_{1d}\right\}  $ when each subsystem can be
considered as being at thermal equilibrium, and the spin-lattice relaxation
process can be ignored. \ 

To derive an equation describing the establishment of equilibrium state (1),
we take the density matrix in the general form \cite{Zubarev1974}%

\begin{equation}
\rho_{ne}\left(  t\right)  =Z_{ne}^{-1}\exp\left(  -\sum_{n}\left(  \beta
_{n}\left(  t\right)  H_{n}-\beta_{n}\left(  t\right)  \int_{-\infty}^{0}d\tau
e^{\varepsilon\tau}K_{n}\left(  \tau\right)  \right)  \right)  \text{, }n=z,dd
\tag{8}%
\end{equation}
where
\begin{equation}
Z_{ne}\left(  t\right)  =Tr\left\{  \exp\left(  -\sum_{n}\left(  \beta
_{n}\left(  t\right)  H_{n}-\beta_{n}\left(  t\right)  \int_{-\infty}^{0}d\tau
e^{\varepsilon\tau}K_{n}\left(  \tau\right)  \right)  \right)  \right\}  ,
\tag{9}%
\end{equation}
$K_{n}\left(  \tau\right)  $ is the operator of the energy flux
\begin{equation}
K_{n}\left(  \tau\right)  =\frac{dH_{n}\left(  \tau\right)  }{d\tau} \tag{10}%
\end{equation}
and transition to the limit $\varepsilon\longrightarrow0$ should be made after
the calculation of the integral. According to the method of the
non-equilibrium statistical operator , the operators under the integral in Eq.
(10) are taken in the Heisenberg representation to be $H_{n}\left(
\tau\right)  =e^{i\tau H}H_{n}e^{-i\tau H}$ \cite{Zubarev1974}.

Accounting that $\sum_{n}K_{n}\left(  \tau\right)  =0$, we rewrite density
matrix (8) in the following form \cite{Zubarev1974}
\begin{equation}
\rho_{ne}=Z_{ne}^{-1}\exp\left(  -\sum_{k}\beta_{k}\left(  t\right)
H_{k}-\left(  \beta_{d}\left(  t\right)  -\beta_{z}\left(  t\right)  \right)
\int_{-\infty}^{0}d\tau e^{\varepsilon\tau}K\left(  \tau\right)  \right)  .
\tag{11}%
\end{equation}

Evolution of the spin system can be sufficiently completely described by the
density matrix (11) with temporally depending $\beta_{z}$ and $\beta_{d}$. To
determine the temporal dependence of the inverse temperatures $\beta
_{z}\left(  t\right)  \ $and $\beta_{d}\left(  t\right)  $, let us calculate
the average energy flux. This flux is given by averaging expression (10) using
non-equilibrium density matrix (11)%

\begin{equation}
\left\langle K_{n}\left(  \tau\right)  \right\rangle =\left\langle
\frac{dH_{n}\left(  \tau\right)  }{d\tau}\right\rangle , \tag{12}%
\end{equation}
where $\left\langle ...\right\rangle =Tr\left(  \rho_{ne}...\right)  $. Eqs.
(11) and (12) are valid at any temperatures of the subsystems and can be
applied to consider dynamics of the dipolar coupled spin system and generation
of the entangled state in the system. The quantities in Eq. (12), averaged
with density matrix (11) , depend nolinearly on the inverse temperatures
$\beta_{z}$ and $\beta_{d}$ and their difference $\beta_{z}-\beta_{d}.$Eq.
(12) is solved using different approximations: the high-temperature
approximation, $\frac{\left\vert H\right\vert }{k_{B}T}\ll1$; the
approximation of small values {}{}of the temperature difference, $\left\vert
\beta_{z}-\beta_{d}\right\vert $ $\ll\min\left\{  \beta_{z},\beta_{d}\right\}
$ or/and the approximation of small energy flux, $\left\Vert K\right\Vert \ll$
$\left\Vert H_{k}\right\Vert $ $\left(  \left\Vert ...\right\Vert \text{
denotes a norm of an operator}\right)  $\cite{Zubarev1974}.

As it is shown below, the entangled states can be achieved in the cases when a
difference $\left(  \beta_{d}-\beta_{z}\right)  $ cannot be regarded as a
small parameter. We cannot also use the high-temperature approximation and the
expansion of (11) in powers of $K$ . Hence, the perturbation theory methods
cannot be applied effectively to study entanglement in the considered systems.

We will solve numerically Eq. (12) for four-spin systems with various
geometry: chain, circle, and rectangle. Examples of such systems are
quasi-one-dimensional fluorine chains in calcium fluorapatite $Ca_{5}F\left(
PO_{4}\right)  _{3}$ \cite{Cho1996} , xenon tetrafluoride with chemical
formula $XeF_{4}$(its crystalline square planar structure was determined by
both NMR spectroscopy and by neutron diffraction studies \cite{Thomas,Burns}),
and 1,4-dichlorobenzene $C_{6}H_{4}Cl$ with a rectangle planar structure.

We use dimensionless units where the dipolar coupling constant of the nearest
spins is chosen to be $D=1.$ The time unit is determined by the DDI energy
between these spins and equals $100$ $\mu s$\thinspace$\allowbreak$. The ratio
of Zeeman energy splitting to the dipolar energy is chosen to be $\frac
{\omega_{0}}{\omega_{d}}=43$ for all considered cases. We assume also that the
angles $\theta_{j,k}=\frac{\pi}{2}$ for all pairs of spins in all cases. In
our numerical calculation we used: a) for a chain all angles $\varphi_{j,k}=$
$\frac{\pi}{2}$ and the coupling constants of spins the $j$-th and $k$-th are
\ $D_{j,k}=\left(  j-k\right)  ^{-3}$; b) for a circle $\varphi_{12}=$
$\frac{\pi}{2}$, $\varphi_{23}=$ $\varphi_{14}=\pi$, $\varphi_{34}=$
$\frac{3\pi}{2}$, $\varphi_{13}=$ $\frac{3\pi}{4}$, $\varphi_{24}=$
$\frac{5\pi}{4}$ \ and $D_{j,k}=\left(  \frac{\sin\frac{\pi}{N}}{\sin\left(
\frac{\pi}{N}\left(  j-k\right)  \right)  }\right)  ^{3}$; and c) for a
rectangle $\varphi_{12}=$ $\frac{\pi}{2}$, $\varphi_{23}=$ $\varphi_{14}=\pi$,
$\varphi_{34}=$ $\frac{3\pi}{2}$, $\varphi_{13}=\arccos\left(  -\frac{\sqrt
{3}}{2}\right)  $, $\varphi_{24}=\arccos\left(  \frac{\sqrt{3}}{2}\right)
+\pi$, \ and $D_{12}=D_{34}=1$, $\ D_{23}=D_{14}=\frac{1}{3\sqrt{3}}$,
$D_{13}=D_{24}=\frac{1}{8}$. Fig. 1 shows evolution of the inverse
temperatures $\beta_{z}\ $and $\beta_{d}$ toward the equilibrium value of the
four-spin systems in the form of a chain, a circle, and a rectangle. \ 

These systems demonstrate qualitatively similar time dependences of $\beta
_{z}\ $and $\beta_{d}$. At a high ratio $\frac{\omega_{0}}{\omega_{d}}=45$ it
is possible to mark out four stages in the $\beta_{d}$ time dependences. These
stages are clearly seen in the $\beta_{d}$ time dependence for a circle spin
system (Fig. 1a, black line). The first stage, $0<t<0.5$ is characterized by a
fast drop of the inverse temperature about twice . During the second stage,
$0.5<t<2$ , the rate of the $\beta_{d}$ drop is decreased, and it is increased
again at the next stage $2<t<2.5$ . The dependence is substantivally
non-exponential. Only at the last stage when {}{}of the temperature difference
is small, the time dependence can be fitted by an exponential laws with the
characteristic time close to $\tau\sim1$. Of course, these stages can be
marked out in the $\beta_{z}$ time dependences. However the variations of the
Zeeman temperatures are very small because the heat capacity of the Zeeman
subsystem is high. The inverse temperature of the Zeeman subsystem $\beta_{z}$
increases less than by 2.5\% (Fig. 1b), while the inverse temperature of the
dipolar subsystem decreases 70 times (Fig. 1a).

With decreasing a ratio of the Zeeman energy splitting to the dipolar energy,
the behave of the time dependence is changed. Only two stages can be marked
out as a ratio $\frac{\omega_{0}}{\omega_{d}}=20$ (Fig. 1a, blue dash-dotted
line). Both stages can be fitted by the exponential laws with characteristic
times, $\tau$: $\tau=0.6$ for the first stage $0<t<0.5$ and $\tau=3.3$ the
second one by . \ So, the decrease of the ratio of the Zeeman energy splitting
to the dipolar energy leads to the increase of the characteristic time of
achieving the equilibrium state.

\section{\textbf{Entanglement at a low Zeeman and dipolar temperatures}}

Several parameters have been proposed for characterizing the entangled state
of a spin system: the von Neumann entropy, entanglement of formation, log
negativity, concurrence of a pair of spins, and etc.
\cite{Amico2008,Horodecki2009,W. K. Wootters1998,G. Vidal2000,G.
Vidal2002,Berry2006}. We will characterize the entangled states by the
concurrence between two, $m$-th and $n$-th, spins which is defined as \cite{W.
K. Wootters1998}%

\begin{equation}
C_{mn}\left(  \beta_{z},\beta_{d}\right)  =\max\left\{  q_{mn}\left(
\beta_{z},\beta_{d}\right)  ,0\right\}  , \tag{14}%
\end{equation}
with $q_{mn}\left(  \beta_{z},\beta_{d}\right)  =\lambda_{mn}^{\left(
1\right)  }\left(  \beta_{z},\beta_{d}\right)  -\lambda_{mn}^{\left(
2\right)  }\left(  \beta_{z},\beta_{d}\right)  -\lambda_{mn}^{\left(
3\right)  }\left(  \beta_{z},\beta_{d}\right)  -\lambda_{mn}^{\left(
4\right)  }\left(  \beta_{z},\beta_{d}\right)  $. Here $\lambda_{mn}^{\left(
k\right)  }\left(  \beta_{z},\beta_{d}\right)  $ $\left(  k=1,2,3,4\right)  $
are the square roots of eigenvalues, in the descending order, of the following
non-Hermitian matrix:
\begin{equation}
R_{mn}\left(  \beta_{z},\beta_{d}\right)  =\rho_{mn}\left(  \beta_{z}%
,\beta_{d}\right)  \left(  \sigma_{y}\otimes\sigma_{y}\right)  \tilde{\rho
}_{mn}\left(  \beta_{z},\beta_{d}\right)  \left(  \sigma_{y}\otimes\sigma
_{y}\right)  , \tag{15}%
\end{equation}
where $\rho_{mn}\left(  \beta_{z},\beta_{d}\right)  $ is the reduced density
matrix. For the $m$-th and $n$-th spins, the reduced density matrix $\rho
_{mn}$ is defined as \ \ $\rho_{mn}=Tr_{mn}\left(  \rho\right)  $ where
$Tr_{mn}\left(  ...\right)  $ denotes the trace over the degrees of freedom
for all spins except the $m$-th and $n$-th spins. In Eq. (15) $\tilde{\rho
}_{mn}$ is the complex conjugation of the reduced density matrix \ $\rho_{mn}$
and $\sigma_{y}$ is the Pauli matrix $\sigma_{y}=%
\begin{pmatrix}
0 & -i\\
i & 0
\end{pmatrix}
$. For maximally entangled states, the concurrence is $C_{mn}=1$ while for
separable states $C_{mn}=0$.

Figs. 2 and 3 present the results of numerical calculations of the concurrence
$C_{mn}$ in the four-spin circle (Figs. 2a and 3a), rectangle (Figs. 2b and
3b), and chain (Figs. 2c and 3c) . Dependences of concurrencies on inverse
temperatures $\beta_{z}$ and $\beta_{d}$ at $\frac{\omega_{0}}{\omega_{d}}=45$
are presented in Figs. 2. Here the dashed red lines present concurrencies in
the equilibrium states ( $\beta_{z}$= $\beta_{d}$) and the dotted yellow lines
show dynamic concurrency variations at achieving the equilibrium state.

\subsection{Conclusion}

We have investigated the entanglement evolution during establishing an
equilibrium in four-spin dipolar coupled systems with various structure:
circle, rectangle, and chain. The concurrence was considered as a measure of
entanglement. Entanglement is observed in non-equilibrium states ( $\beta_{z}$
$\neq\beta_{d}$) and can appear when both temperatures much higher than the
temperature of the entanglement appearance in the equilibrium state. This
confirms the results of previous works that non-equilibrium conditions are
more favorable for the generation of entangled states as equilibrium states
\cite{Galve2010,V. Vedral2010}. Entangled states appear and are stored at
lower dipolar temperatures relative to the Zeeman one$\left(  \ \beta_{d}%
^{-1}<\beta_{z}^{-1}\right)  $. We found that the concurrence
non-monotonically depends on both temperatures (Fig. 2).

Considering the establishment of the equilibrium state, we showed that the
well-known high-temperature approximation is not applicable for a homonuclear
spin system at low temperature, and the temporal dependence of the inverse
temperatures of the spin subsystems is given by a solution of non-linear equations.

We demonstrated that the concurrence dependences on time is non-monotonic
during the equilibrium state establishment process (Fig. 3) and entanglement
fades long before equilibrium is established (compare Figs. 1 and 3). The
behavior of entanglement depends strongly on the ratio of the Zeeman energy to
the dipolar energy. At high ratio, the concurrence in the system decreases
significantly till the residual very small quantity in time of the order of
$T_{2}$. At low ratio, the establishment of equilibrium (Fig. 1a and b, blue
dash-dotted line) and the decay of entanglement (Fig. 3a, green dotted line)
take time up to 1 $ms$.

Obtained results open an effective way to the experimental testing of
entanglement in spin systems by using non-equilibrium states to generate the
entangled state at higher temperatures.

Figure Captions

\bigskip

Fig. 1 Equalization of the inverse spin temperatures of the dipolar (a) and
Zeeman (b) subsystems in a four spin circle (solid black line), chain (dashed
red line), and rectangle (dotted yellow line) at $\frac{\omega_{0}}{\omega
_{d}}=45$. Blue dash-dotted line -- the inverse spin temperatures of the
dipolar (a) and Zeeman (b) subsystems in a four spin circle at $\frac
{\omega_{0}}{\omega_{d}}=20$ .

\bigskip

Fig. 2 The concurrence in four spin system as a function of the inverse
temperatures $\beta_{z}$ and $\beta_{d}$: (a) $C_{12}$ in the circle, (b)
$C_{13}$ in the rectangle , and (c) $C_{14}$ in the chain. Dashed red line -
equilibrium state with $\beta_{z}$= $\beta_{d}$. Dotted yellow line --
evolution of the concurrence under equalization of the Zeeman $\beta_{z}$ and
$\beta_{d}$ inverse temperatures due the influence of the non-secular term
$H_{nd}$.

\bigskip

Fig. 3 Time dependence of concurrence $C_{mn}$ in a four spin system at
$\frac{\omega_{0}}{\omega_{d}}=45$: (a) circle, (b) rectangle, (c) chain.
Solid black line -- $C_{12}$, dashed red line -- $C_{13}$, and dotted blue
line -- $C_{14}$. Green dotted line -- $C_{12}$ in a four circle at
$\frac{\omega_{0}}{\omega_{d}}=20$.

\end{document}